\documentclass[prl,twocolumn,showpacs,amsmath,amssymb]{revtex4}
\usepackage[dvips]{epsfig}

\begin{document}

\preprint{YARU-HE-02/03 \\ hep-ph/0204201}

\title{Electron Mass Operator in a Strong Magnetic Field \\
and Dynamical Chiral Symmetry Breaking}

\author{A. V. Kuznetsov}
 \email{avkuzn@uniyar.ac.ru}
\author{N. V. Mikheev}
 \email{mikheev@uniyar.ac.ru}
\affiliation{
Division of Theoretical Physics, Department of Physics,\\
Yaroslavl State University, Sovietskaya 14,\\
150000 Yaroslavl, Russian Federation}

\date{April 2, 2002}

\begin{abstract}
The electron mass operator in a strong magnetic field is calculated. 
The contribution of higher Landau levels of virtual electrons, along 
with the ground Landau level, is shown to be essential in the leading 
log approximation.
The effect of the electron dynamical mass generation by a magnetic field 
is investigated.
In a model with $N$ charged fermions, it is shown that some critical 
number $N_{cr}$ exists for any value of the electromagnetic coupling 
constant $\alpha$, such that the fermion dynamical mass is generated 
with a doublet splitting for $N < N_{cr}$, and the dynamical mass 
does not arise at all for $N > N_{cr}$, thus leaving the chiral symmetry 
unbroken.
\end{abstract}

\pacs{11.30.Rd, 11.30.Qc, 12.20.Ds}

\maketitle

Asymptotic properties of the QED diagrams and operators 
in strong magnetic fields
$B \gg B_e, \,B_e = m_e^2/e \simeq 4.41 \cdot 10^{13}\,\text{G}$ 
($e$ is the elementary charge) 
are of conceptual interest both from the standpoint of the searches 
of the perturbation theory applicability borders, and also 
in view of possible applications in astrophysics and in cosmology of the 
early Universe.
The investigations of this type are being performed by many authors during 
a rather long time. For example, a history of calculations of 
the electron mass operator in a strong magnetic field lasts more than 
30 years already. However, as we show in this Letter, it is too early 
to put the final point in the problem.

One-loop contribution into the electron mass operator in a strong 
magnetic field was obtained for the first time by 
Jancovici~\cite{Jancovici:1969} in the leading (double) log approximation. 
Later on, in the papers by Loskutov and 
Skobelev~\cite{Loskutov:1979,Loskutov:1981} the attempts were performed to 
calculate the two-loop contribution and to summarize all the many-loop 
contributions in the same approximation.  
A correct formula for the electron mass operator in this 
approximation was obtained recently in the paper~\cite{Gusynin:1999a}. 
However, the double log approximation becomes invalid in 
asymptotically strong magnetic fields~\cite{Loskutov:1983}, 
because of the crucial influence of the strong magnetic field 
on the virtual photon polarization operator. 
This influence provides an appearance of the effective photon mass, 
$m_\gamma^2 = (2\alpha/\pi) e B$, which replaces the electron mass in 
one of the two logarithms. A correct expression for the electron 
mass operator in the leading (single) log approximation was obtained 
recently in our paper~\cite{Kuznetsov:2002} by the summation of 
the rainbow Feynman diagrams, in the form:
\begin{equation}
M = \frac{m_0}
{1 - (\alpha_R/2 \pi) \, [\ln (\pi/\alpha_R) - \gamma_{\text{E}}] \,
\ln (e B/m_0^2)},
\label{eq:M_our}
\end{equation}
where $m_0$ is the electron mass without field,
$\gamma_{\text{E}} = 0.577 \dots$ is the Euler constant,
$\alpha_R$ is the electromagnetic coupling constant renormalized by 
the field
\begin{equation}
\alpha_R = 
\frac{\alpha}{1 - (\alpha/3 \pi) \; \ln (e B/m_0^2)}. 
\label{eq:alphaR}
\end{equation}
However, the formula~(\ref{eq:M_our}) has a restricted area 
of applicability with respect to the parameter 
$\eta_0 = \frac{\alpha}{2 \pi} \, \ln \frac{e B}{m_0^2}$, 
because the mass operator~(\ref{eq:M_our}) tends to infinity at some 
finite value of this parameter. 
On the one hand, as it was mentioned in~\cite{Gusynin:1999a}, this could 
be the signal of a new physics, the spontaneous breaking of chiral symmetry. 
On the other hand, it indicates that the formula~(\ref{eq:M_our}) is not 
self-consistent at the field values for which $\eta_0 \sim 1$.  
It means that the transitions to the nonperturbative regime over the 
parameter $\eta_0$ is necessary. In terms of the diagram technique it 
leads to the necessity to consider, along with the rainbow diagrams 
giving the leading log contributions, also the diagrams 
of a next-to-leading order.
As the analysis shows, this extension reduces to a substitution of the
electron mass operator instead of mass into the diagrams. 
It is more convenient, however, to use the Schwinger-Dyson equations 
as it was performed in 
Refs.~\cite{Gusynin:1999b,Gusynin:1999c,Alexandre:2000,Alexandre:2001}.

\begin{figure}[htb]
\includegraphics{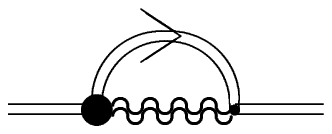}
\caption{\label{fig:ESED}
Feynman diagram for the field-induced contribution 
into the electron mass operator. Double lines correspond 
to exact solutions and exact propagators of electrons and 
photons in an external magnetic field. 
Bold circle depicts the exact vertex.}
\end{figure}

\begin{figure}[htb]
\includegraphics{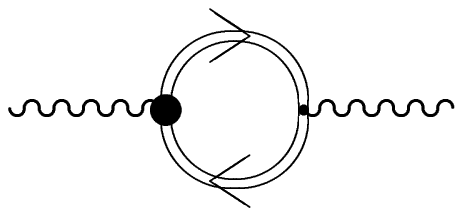}
\caption{\label{fig:PSED}
Feynman diagram for the field-induced contribution 
into the photon polarization operator.}
\end{figure}

In this Letter, we realize this program by calculating the electron 
mass operator to be valid in any asymptotically strong magnetic fields. 
To obtain a full system of the Schwinger-Dyson equations in the presence 
of external magnetic field, 
which corresponds to the irreducible diagrams depicted in Figs.~\ref{fig:ESED} 
and~\ref{fig:PSED}, one should know the exact vertex $\Gamma_\mu$ which 
contains, as is well-known, an infinite number of irreducible diagrams. 

Fortunately, the problem is simplified essentially in the strong field 
limit, as it was shown for the first time in Ref.~\cite{Loskutov:1984}, 
see also~\cite{Gusynin:1999b,Gusynin:1999c}. Namely, the exact vertex 
$\Gamma_\mu$ can be reduced to the bare vertex $\gamma_\mu$
in an appropriate gauge. 

In the strong field limit, the main contribution into the electron 
mass operator originates from virtual electrons occupying the lowest 
Landau level~\cite{Loskutov:1976}. In this case the Fourier transform 
of the translational invariant part of the exact electron propagator 
in a magnetic field can be presented in the form
\begin{equation}
G_0^{(e)} (k) = i \exp \left(- \frac{k_{\bot}^2}{2 e B} \right) \, 
\left[\hat k_{\|} - \hat M (k_{\|})\right]^{-1} \, O^{(-)},
\label{eq:LLL_prop}
\end{equation}
where $\hat k_{\|} = k^0 \gamma^0 - k^3 \gamma^3$ (the field is 
directed along the 3d axis), 
$\hat M (k_{\|})$ is the sought mass operator,
$O^{(-)} = (1 - i \gamma^1 \gamma^2)/2$ is the projecting operator 
corresponding to the electron state with the spin antiparallel 
to the external field direction.
 
On the other hand, photons of the only one mode ('transversal' in the 
Adler's notation~\cite{Adler:1971}) interact with electrons occupying 
the lowest Landau level. In this case, in the gauge where 
$\Gamma_\mu \simeq \gamma_\mu$ the exact photon propagator can be presented 
in the form~\cite{Loskutov:1983} 
\begin{eqnarray}
&&G^{(\gamma)}_{\mu \nu} (q) = - i \; {\mathcal D} (q_{\bot}^2,q_{\|}^2) \;
\tilde\Lambda_{\mu \nu}, 
\label{eq:G_munu}\\
&&{\mathcal D} (q_{\bot}^2,q_{\|}^2) 
= \frac{1}{q^2 - {\mathcal P} (q_{\bot}^2,q_{\|}^2)} .
\label{eq:D_q}
\end{eqnarray}
Here $q_{\bot}^2 = q_\mu \Lambda_{\mu \nu} q_\nu$, 
$q_{\|}^2 = q_\mu \tilde\Lambda_{\mu \nu} q_\nu$, 
$q^2 = q_{\|}^2 - q_{\bot}^2$,
$\Lambda_{\mu \nu} = \varphi_{\mu \rho} \varphi_{\rho \nu}$, 
$\tilde\Lambda_{\mu \nu} = \tilde\varphi_{\mu \rho} \tilde\varphi_{\rho \nu}$, 
$\varphi_{\alpha \beta}$ is the dimensionless tensor of external 
magnetic field,
$\varphi_{\alpha \beta} = F_{\alpha \beta} \left / 
\sqrt{F_{\mu \nu}^2 / 2} \right .$, 
$\tilde\varphi_{\alpha \beta} = \frac{1}{2} 
\varepsilon_{\alpha \beta \rho \sigma} 
\varphi_{\rho \sigma}$ is the dual tensor.
The function ${\mathcal P} (q_{\bot}^2,q_{\|}^2)$ is the eigenvalue 
of the photon polarization operator ${\mathcal P}_{\mu \nu} (q)$, 
which is depicted by the Feynman diagram Fig.~\ref{fig:PSED}. 
With the reduction $\Gamma_\mu = \gamma_\mu$, the operator 
has the form
\begin{eqnarray}
{\mathcal P}_{\mu \nu} (q) &=& - i \frac{\alpha}{4 \pi^3} \,
\int d^4 k \, Tr \left[\gamma_\mu G^{(e)} (k) \gamma_\nu G^{(e)} (k-q) 
\right] 
\nonumber\\
&=& 
\left(\tilde\Lambda_{\mu \nu} - 
\frac{q_{\mu \|} q_{\nu \|}}{q_{\|}^2} \right) 
\, {\mathcal P} (q_{\bot}^2,q_{\|}^2) + \dots , 
\label{eq:P_munu}
\end{eqnarray}
where dots denote the contribution of the other photon modes.
Thus the polarization operator~(\ref{eq:P_munu}) is reduced in fact 
to the one-loop operator. 

The mass operator $\hat M (p_{\|})$ corresponding to the irreducible 
diagram depicted in Fig.~\ref{fig:ESED}, in the gauge 
$\Gamma_\mu \simeq \gamma_\mu$ and in terms of Eqs.~(\ref{eq:LLL_prop}) 
and~(\ref{eq:G_munu}) is reduced to a function
$M (p_{\|}^2)$, see Refs.~\cite{Gusynin:1999b,Gusynin:1999c}. 
The following integral equation arises for this function 
\begin{widetext}
\begin{equation} 
M (p_{\|}^2) = m_0 - i \frac{\alpha}{2 \pi^3} \,
\int d^4 k \, \exp \left(- \frac{k_{\bot}^2}{2 e B} \right) \,
\frac{M (k_{\|}^2)}{k_{\|}^2 - [M (k_{\|}^2)]^2} \,
\frac{1}{(k-p)_{\|}^2 - k_{\bot}^2 -
{\mathcal P} \left( k_{\bot}^2,(k-p)_{\|}^2 \right)}.
\label{eq:M_fun}
\end{equation}
\end{widetext}
Integral in the expression~(\ref{eq:M_fun}) does not contain
an ultraviolet divergency, because the integration over the 
momenta $k_{\bot}$ transversal to the field direction has the cutoff 
$k_{\bot} \sim \sqrt{e B}$. 

On the other hand, the photon polarization operator~(\ref{eq:P_munu}), 
in general,  
does contain the ultraviolet divergency. As a result, virtual electrons 
occupying both the lowest and higher Landau levels contribute to the 
integral. This fact was not taken into account in all previous 
publications in the field. However, as will be shown below, it leads 
to very interesting physical consequences.

The function ${\mathcal P} (q_{\bot}^2,q_{\|}^2)$ in a strong 
magnetic field in the one-loop approximation can be extracted, for example, 
from Ref.~\cite{Tsai:1974}, see also~\cite{Shabad:1992vx}, 
where the sought function $M (q_{\|}^2)$ 
should be inserted instead of the field-free mass $m_0$: 
\begin{eqnarray}
{\mathcal P} (q_{\bot}^2,q_{\|}^2) &=& - \frac{2 \alpha}{\pi} \; e B \; 
\exp \left(- \frac{q_{\bot}^2}{2 e B} \right) \, 
H\!\! \left( \frac{q_{\|}^2}{4 [M (q_{\|}^2)]^2} \right) 
\nonumber\\*
&+& \frac{\alpha}{3 \pi} \; q^2 \; \ln \frac{e B}{[M (q_{\|}^2)]^2}.  
\label{eq:P_q^2}
\end{eqnarray}
Here the first term is caused by virtual electrons occupying the lowest 
Landau level, while the second term contains the contribution from 
higher Landau levels. The function $H(z)$ has the form
\begin{equation}
H(z) = \frac{1}{2 \sqrt{-z(1 - z)}} \ln 
\frac{\sqrt{1 - z} + \sqrt{-z}}{\sqrt{1 - z} - \sqrt{-z}} - 1. 
\label{eq:H(z)}
\end{equation}
In the analysis of Eq.~(\ref{eq:M_fun}), we are interested in the region 
of parameters $q_{\|}^2 < 0$, $|q_{\|}^2| \gg [M (q_{\|}^2)]^2$.  
For large negative values of the argument, the function $H(z)$ is 
simplified, $H(z) \simeq -1$. The first term in Eq.~(\ref{eq:P_q^2}) 
acquires in this case the meaning of the photon mass squared,
$m_\gamma^2 = (2 \alpha/\pi) e B$, induced by a magnetic field.
As for the second term in Eq.~(\ref{eq:P_q^2}), containing the 
contribution from higher Landau levels into the photon polarization 
operator, its role is in renormalization of the electromagnetic 
constant $\alpha$ in a magnetic field, $\alpha \to \alpha_R$. 
The expression for the renormalized constant can be obtained from 
Eq.~(\ref{eq:alphaR}) by the replacement $m_0 \to M (q_{\|}^2)$. 

Turning back to Eq.~(\ref{eq:M_fun}), it should be mentioned that 
the integral in the right-hand side exactly corresponds to the one-loop 
field-induced correction to the electron mass, with the replacement 
of the field-free mass $m_0$ by the sought mass operator $M (k_{\|}^2)$ 
in the integrand. In the superstrong field limit, the main contribution 
into the integral~(\ref{eq:M_fun}) in the form of a big logarithm 
$\ln (e B/M^2)$ originates from the region
$[M (k_{\|}^2)]^2 \ll |k_{\|}^2| \ll m_\gamma^2 \sim \alpha e B$, 
$m_\gamma^2 \alt k_{\bot}^2 \ll e B$. 
In view of this, the calculation of the integral in Eq.~(\ref{eq:M_fun}) 
with logarithmic accuracy gives the following result for the mass 
operator $M (p_{\|}^2)$ in the Euclidean region
$p_{\|}^2 < 0, \, |p_{\|}^2| \ll m_\gamma^2$: 
\begin{equation}
M (p_{\|}^2) \simeq M (0) \left[1 + \frac{p_{\|}^2}{4 e B}\,
\ln \frac{2 e B}{[M (0)]^2}\right].
\label{eq:M_res}
\end{equation}
The expression~(\ref{eq:M_res}) shows that the electron physical mass 
which is defined, strictly speaking, as the solution of the dispersion 
equation $m = M (-m^2)$, can be taken with a great accuracy in the zero 
point, $m \simeq M (0)$. 

In the leading log approximation, we have obtained the following 
transcendental equation for the electron physical mass from the 
integral equation~(\ref{eq:M_fun}): 
\begin{equation}
m = m_0 + m \, \frac{\alpha_R}{2 \pi} \, 
\left(\ln \frac{\pi}{\alpha_R} - \gamma_{\text{E}} \right) \,
\ln \frac{e B}{m^2},
\label{eq:m_eq}
\end{equation}
where $\alpha_R$ is taken 
in the point $q_{\|}^2 = 0$, i.e. obtained from Eq.~(\ref{eq:alphaR}) 
by the replacement $m_0 \to M (0) = m$. 
It is interesting to note that the formula~(\ref{eq:m_eq}) reproduces 
exactly our result~(\ref{eq:M_our}), with the substitution of the 
electron physical mass $m$ instead of the field-free mass $m_0$ under 
the logarithms. 
The equation~(\ref{eq:m_eq}) solves the problem of finding the electron 
physical mass for any large values of the magnetic field. 
It is free of a singularity, unlike the Eq.~(\ref{eq:M_our}).

An analysis of Eq.~(\ref{eq:m_eq}) shows that its solution 
in asymptotically strong fields when $m \gg m_0$, becomes independent 
on $m_0$ and is reduced in fact to the solution at $m_0 = 0$. 
This would mean the generation of the dynamical mass of the initially 
massless electron in a magnetic field. 
This effect which is also called the dynamical chiral symmetry breaking, 
was studied in refs.~\cite{Gusynin:1999a,Gusynin:1999b,Gusynin:1999c,%
Alexandre:2000,Alexandre:2001}, however, the contribution from higher 
Landau levels into the photon polarization operator was not considered 
there.

Let us show, that this contribution changes the behaviour
of the dynamical mass essentially. Let us extend our analysis to a model 
with $N$ initially massless fermions ($m_0 = 0$) with equal charges $e$ 
(in a case of different fermion charges $Q_f e$, the parameter 
$N$ has the meaning of $\sum_f Q_f^2$). 
In this case the photon polarization operator~(\ref{eq:P_q^2}) is the sum 
over all fermion loops, i.e. it should be multiplied by $N$. 
The transcendental equation for the fermion dynamical mass ($m \ne 0$) 
can be obtained from Eq.~(\ref{eq:m_eq}) as follows
\begin{equation}
\frac{\alpha_R}{2 \pi} \, 
\left(\ln \frac{\pi}{N \alpha_R} - \gamma_{\text{E}} \right) \,
\ln \frac{e B}{m^2} = 1,
\label{eq:m_eq_N}
\end{equation}
where
\begin{equation}
\alpha_R = 
\frac{\alpha}{1 - (N \alpha/3 \pi) \; \ln (e B/m^2)}. 
\label{eq:alphaR_N}
\end{equation}
The expression~(\ref{eq:m_eq_N}) allows to reproduce 
the result of Refs.~\cite{Gusynin:1999b,Gusynin:1999c}, 
if the actual dependence of the coupling constant $\alpha_R$ on the ratio 
$e B/m^2$ is formally ignored and $\alpha_R = \alpha$ is taken in it. 
It is remarkable that the constant $C_1$ obtained there by a numerical 
calculation as $C_1 \simeq 1.82 \pm 0.06$, appears to be 
$C_1 = \pi \exp (- \gamma_{\text{E}}) = 1.763877 \dots$.

\begin{figure}[htb]
\psfig{file=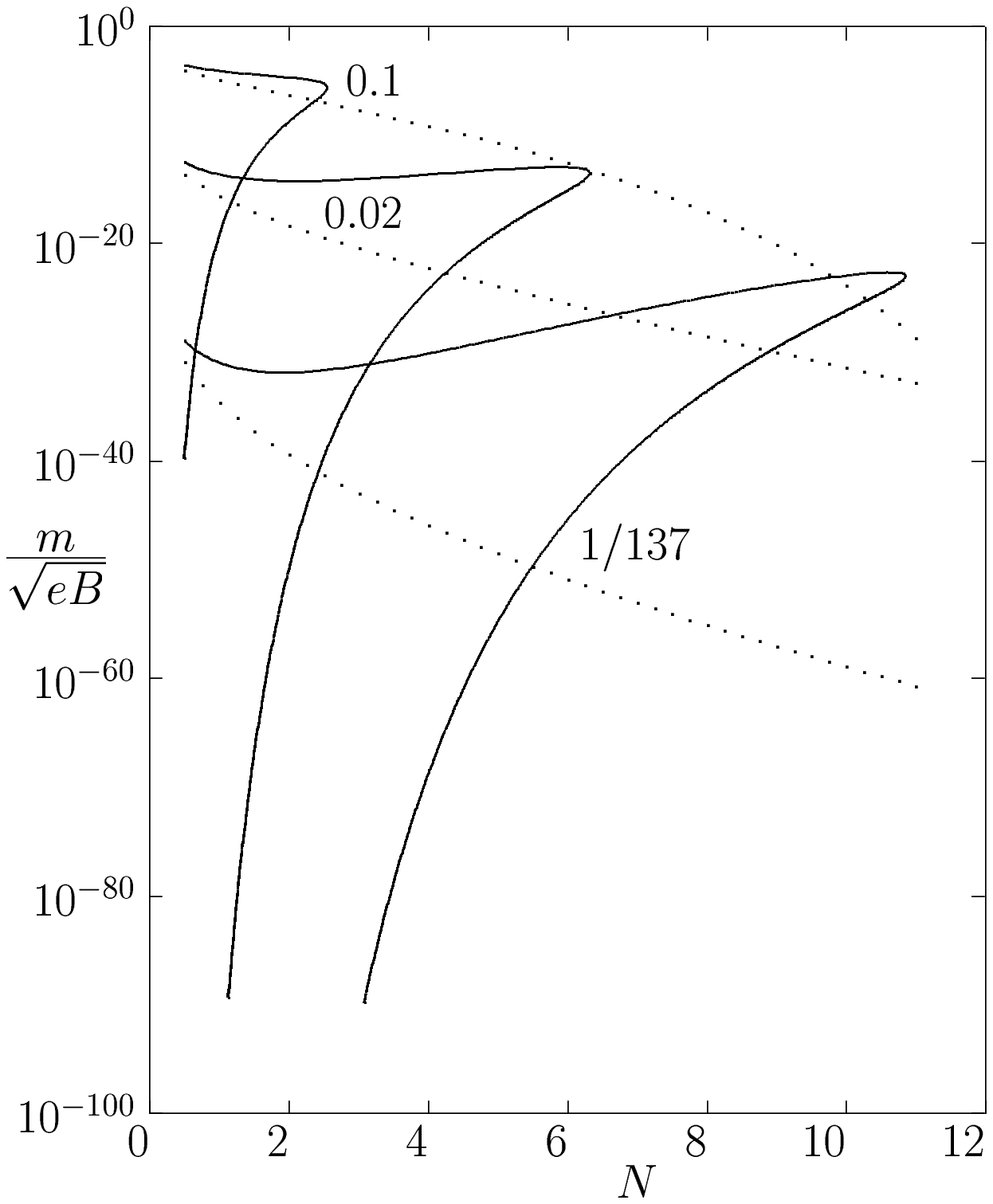,width=0.4\textwidth} 
\caption{\label{fig:m_dyn}
The dependence of the fermion dynamical mass $m$ divided by
$\sqrt{e B}$ on the number of fermions $N$ for different values of 
the field-free coupling constant, $\alpha = 0.1, \, 0.02, \, 1/137$.
Solid lines correspond to our result Eq.~(\ref{eq:m_eq_N}). 
The dotted lines present the results of 
Refs.~\cite{Gusynin:1999b,Gusynin:1999c}.}  
\end{figure}

In Fig.~\ref{fig:m_dyn} the behaviour of the fermion dynamical mass $m$ 
divided by $\sqrt{e B}$ is shown versus the number of fermions $N$ 
and the field-free coupling constant $\alpha$, considered as free 
parameters of the model.

The dependence is seen to differ essentially from the results of 
Refs.~\cite{Gusynin:1999b,Gusynin:1999c}. Namely, for any value of 
the coupling constant $\alpha$, such a critical number of fermions $N_{cr}$
exists that for $N < N_{cr}$ two values of the fermion dynamical mass are 
generated. For $N > N_{cr}$ the equation~(\ref{eq:m_eq_N}) does not have 
a solution at all, thus the chiral symmetry is kept unbroken. 

The dependence of the critical number $N_{cr}$ on the value of 
the coupling constant $\alpha$ takes the form
\begin{eqnarray}
N_{cr} (\alpha) &=& \sqrt{\frac{3 \pi}{2 \exp(\gamma_{\text{E}} + 1)}} \,
\frac{1}{\sqrt{\alpha}} - \frac{3}{4} 
\nonumber\\*
&+& \frac{9}{16} \,
\sqrt{\frac{3 \exp(\gamma_{\text{E}} + 1)}{2 \pi}} \, \sqrt{\alpha} 
+ O (\alpha).
\label{eq:N_cr}
\end{eqnarray}

It is quite remarkable that the doublet splitting of the fermion 
dynamical mass can be rather large. For example, for $\alpha = 0.1$ and $N=1$ 
the mass difference is of 15 orders of magnitude. If one considers for 
the purposes of illustration the magnetic field value 
$\sim 10^{33}\,\text{G}$ \cite{Ambjorn:1993}, the two values of the fermion 
dynamical mass are $\sim 10^2\,\text{GeV}$ and $\sim 10^{-4}\,\text{eV}$.

We believe that the doublet splitting of the fermion 
dynamical mass and the conservation of the chiral symmetry 
at $N > N_{cr}$ as well are new interesting physical phenomena in QED 
in strong external magnetic field. 

\begin{acknowledgments}
We are grateful to V.A. Rubakov and M. V. Chistyakov for helpful discussions.

This work was supported in part by the Russian Foundation for Basic 
Research under the Grant No. 01-02-17334
and by the Ministry of Education of Russian Federation under the 
Grant No. E00-11.0-5.
\end{acknowledgments}

\newpage
\bibliography{kuzn}

\end{document}